\documentclass[aps,prl,twocolumn,showpacs,reprint,nofootinbib]{revtex4-1}
\usepackage{bm,hyperref}
\usepackage{amsmath}
\usepackage{graphicx}
\usepackage{soul}
\usepackage{amsfonts}
\usepackage{amssymb}
\usepackage{dcolumn}
\usepackage{amsmath}
\usepackage{cancel}
\usepackage{overpic}
\usepackage{float}

\newcommand\Tr[1]{\textrm{Tr}\left\{#1\right\}}

\begin{document}
\title{Core-shell particles as building blocks for systems with high duality symmetry}
\author{Aso Rahimzadegan}
\email{aso.rahimzadegan@kit.edu}
\affiliation{Institut f\"ur Theoretische Festk\"orperphysik, Karlsruhe Institute of Technology, 76131 Karlsruhe, Germany}
\author{Carsten Rockstuhl}
\affiliation{Institut f\"ur Theoretische Festk\"orperphysik, Karlsruhe Institute of Technology, 76131 Karlsruhe, Germany}
\affiliation{Institute of Nanotechnology, Karlsruhe Institute of Technology, 76021 Karlsruhe, Germany}
\author{Ivan Fernandez-Corbaton}
\email{ivan.fernandez-corbaton@kit.edu}
\affiliation{Institute of Nanotechnology, Karlsruhe Institute of Technology, 76021 Karlsruhe, Germany}

\begin{abstract}
Material electromagnetic duality symmetry requires a system to have equal electric and magnetic responses. Electromagnetic duality enables technologically important effects like artificial optical activity and zero back-scattering, is a requirement for metamaterials in transformation optics, Huygens wave-front control, and maximal electromagnetic chirality, and appears in topological photonic systems. Intrinsically dual materials that meet the duality conditions at the level of the constitutive relations do not exist in many frequency bands. Nevertheless, discrete objects like metallic helices and homogeneous dielectric spheres can be engineered to approximate the dual behavior. The discrete objects can then be used as building blocks with the objective of obtaining composite systems with high duality symmetry. Here, we exploit the extra degrees of freedom of a core-shell dielectric sphere to obtain a particle whose duality symmetry is more than one order of magnitude better than previously reported non-magnetic objects. We show that the improvement is transferred onto the duality symmetry of composite objects when the core-shell particle is used as a building block instead of homogeneous spheres.
\end{abstract}
\maketitle
\begin{figure}
\begin{centering}
	\begin{overpic}[clip, trim=5cm 0 1.2cm 0, width=0.48\textwidth]{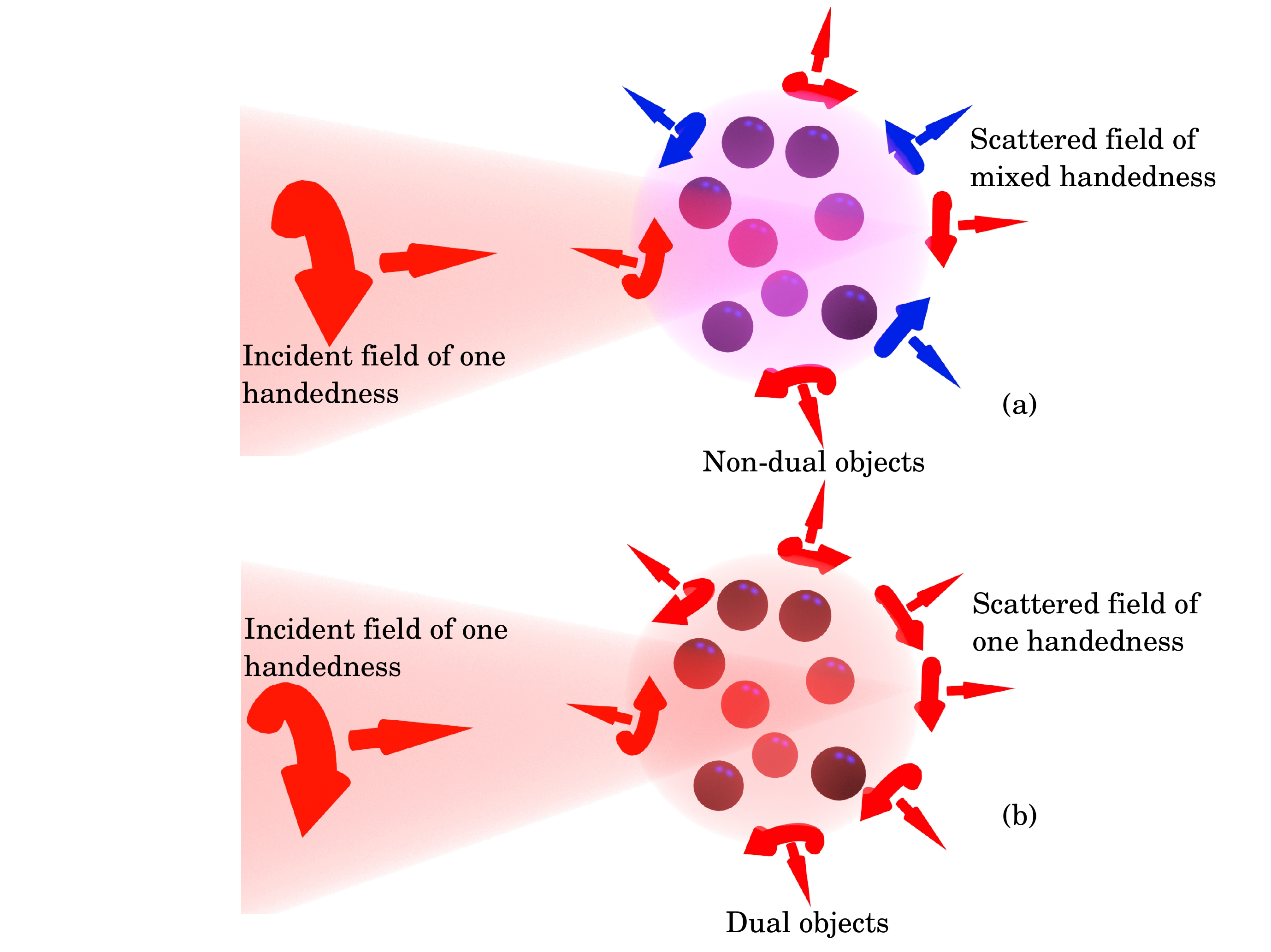}
	\end{overpic}
\par\end{centering}
\protect\caption{An incident field of one handedness illuminates two different systems. The system in (a) is an ensemble of non-dual particles, and the scattered field contains both handedness. The system in (b) is an ensemble of dual particles, and the scattered field contains only the incident handedness.\label{fig: art}}
\end{figure}

A material system that responds in the same way to electric and magnetic fields is said to have electromagnetic duality symmetry, to be invariant under duality, or, for short, to be dual. In much the same way that a translationally invariant system prevents the coupling of plane waves with different momenta during the light-matter interaction, a dual system prevents the coupling of the two helicity components of the field (generalized polarization handedness), as illustrated in Fig. \ref{fig: art}. This statement is valid for all fields: Evanescent, propagating, collimated, focused, far, near, etc ... . Studies show that dual objects enable many technologically important effects like zero back-scattering \cite{Wagner1963,Kerker1983,Karilainen2012,Zambrana2013,FerCor2013c, Zhang2017}, generalized optical activity \cite{Fernandez-Corbaton2015}, transformation media \cite{Leonhardt2009,Thompson2010,Thompson2011,FerCor2013}, wave-front control \cite{Pfeiffer2013,Chong2016,Ziolkowski2017}, chiral transparency \cite{FerCor2016,rahimzadegan2016optical}, the coherent control of light \cite{Schmidt2015}, and appear in systems for the topologically protected propagation of edge states \cite[Fig. 3]{Khanikaev2013}\cite[Sec. III]{Silveirinha2017}.

The conditions for a material to be intrinsically dual at the level of its constitutive relations \cite{Lindell2009,FerCor2013}, roughly speaking an equal electric and magnetic response, are not met by any known natural material at optical frequencies. This rules out the possibility to fashion dual objects in this way. On the other hand, discrete objects made of intrinsically non-dual materials can be engineered to be approximately dual. Duality breaking can be understood as the degree with which the system couples the two helicities of the field, and its quantification can be made using the T-matrix of the system, as we describe in detail below. Duality breakings of 1 part in 10 for small metal helices \cite{FerCor2016}, 1 part in 100 for large homogeneous dielectric spheres \cite{Abdelrahman2017}, and 1 part in a 1000 for small homogeneous dielectric spheres \cite{Fernandez-Corbaton2015}, have been reported so far. Using such discrete objects as building blocks for larger composite objects, metasurfaces, or bulk metamaterials is a logical step, but raises an issue. While an arrangement of perfectly dual objects remains perfectly dual [see Fig. \ref{fig: art}(b)], the duality breaking of an arrangement of approximately dual objects is a question for study. For example, the aforementioned small dielectric spheres designed to approximately meet the dipolar duality condition $a_1\approx b_1$, also feature duality breaking higher order Mie coefficients, e.g. $a_2\neq b_2$ \cite{Zambrana2013b}. These higher order terms become more relevant in the near field mediated couplings of closely packed spheres, and some level of duality degradation is to be expected when going from the individual object to a composite one. Degraded duality results in the degradation of the sought after effect, and should hence be minimized. For example, it increases the back-scattering \cite{Abdelrahman2017}, and, in artificial optical activity, it makes the polarization rotation angle increasingly dependent on the input polarization \cite{Fernandez-Corbaton2015}. The availability of particles that deviate as little as possible from the duality condition is therefore important.

In this Letter, we first study the degradation of the duality symmetry of systems composed of closely packed small homogeneous dielectric spheres. We use the spheres that have the lowest duality breaking reported so far. We show that duality symmetry degrades rapidly when two approximately dual homogeneous dielectric spheres are brought closer than 1/10th of the operating wavelength. We then show that the situation is significantly improved by exploiting the extra degrees of freedom of a core-shell particle, which achieves an order of magnitude improvement in the duality of both the individual particles and the dimers made out of them. Finally, we analyze the duality breaking of an ensemble of eight randomly arranged particles ( like in Fig. \ref{fig: art}). The results show the same order of magnitude improvement in duality when the core-shell particle is used with respect to ensembles of homogeneous spheres. 

We start with the definition of a quantitative measure of duality breaking based on the T-matrix of an object. The T-matrix $\mathbf{T}$ contains all the information about the interaction of the object with the electromagnetic field. It is very often expressed in the basis of multipolar fields of well-defined parity \cite{Mishchenko2016}. Then, the T-matrix relates the multipolar coefficients of the incident field $[\mathbf{p};\mathbf{q}]$ to the multipolar coefficients of the scattered field $[\mathbf{a};\mathbf{b}]$:
\begin{equation}
	\label{eq:T}
\left(\begin{array}{c}
\mathbf{a}\\
\mathbf{b}
\end{array}\right)	=	\mathbf{T}\left(\begin{array}{c}
\mathbf{p}\\
\mathbf{q}
\end{array}\right)=\left(\begin{array}{cc}
		\mathbf{T}_{\text{ee}} & \mathbf{T}_{\text{em}}\\
		\mathbf{T}_{\text{me}} & \mathbf{T}_{\text{ee}}
\end{array}\right)\left(\begin{array}{c}
\mathbf{p}\\
\mathbf{q}
\end{array}\right) \\ ,
 \end{equation}
where $\mathbf{a}$ and $\mathbf{p}$ contain electric multipole coefficients and $\mathbf{b}$ and $\mathbf{q}$ magnetic ones. The submatrix $\mathbf{T}_{\text{ee}}$ relates electric coefficients to electric ones, $\mathbf{T}_{\text{me}}$ electric to magnetic, etc. For an isotropic particle, the T-matrix in this basis simplifies to a diagonal matrix whose elements corresponds to the Mie coefficients.
\begin{table*}
\protect\caption{Design parameters and relevant electromagnetic properties of the individual particles embedded in free space ($\lambda$=1$\mu$m). The Mie coefficients are given up to the octupoles.}  \label{table:Force} 

\begin{centering}
\begin{tabular}{|c|c|c|c|c|c|c|c|}
\hline
Particle & Name & Parameters & %
\begin{tabular}{c}
\vspace*{-0.3cm}
\tabularnewline
 $\cancel{D}$\tabularnewline
\vspace*{-0.3cm}
\tabularnewline
\end{tabular} & $C$ & Mie $a_1$,$b_1$ & Mie $a_2$,$b_2$ & Mie $a_3$,$b_3$  \tabularnewline
\hline
\hline
 \includegraphics[trim={1cm 1cm 0.5cm -25cm},clip,scale=0.009]{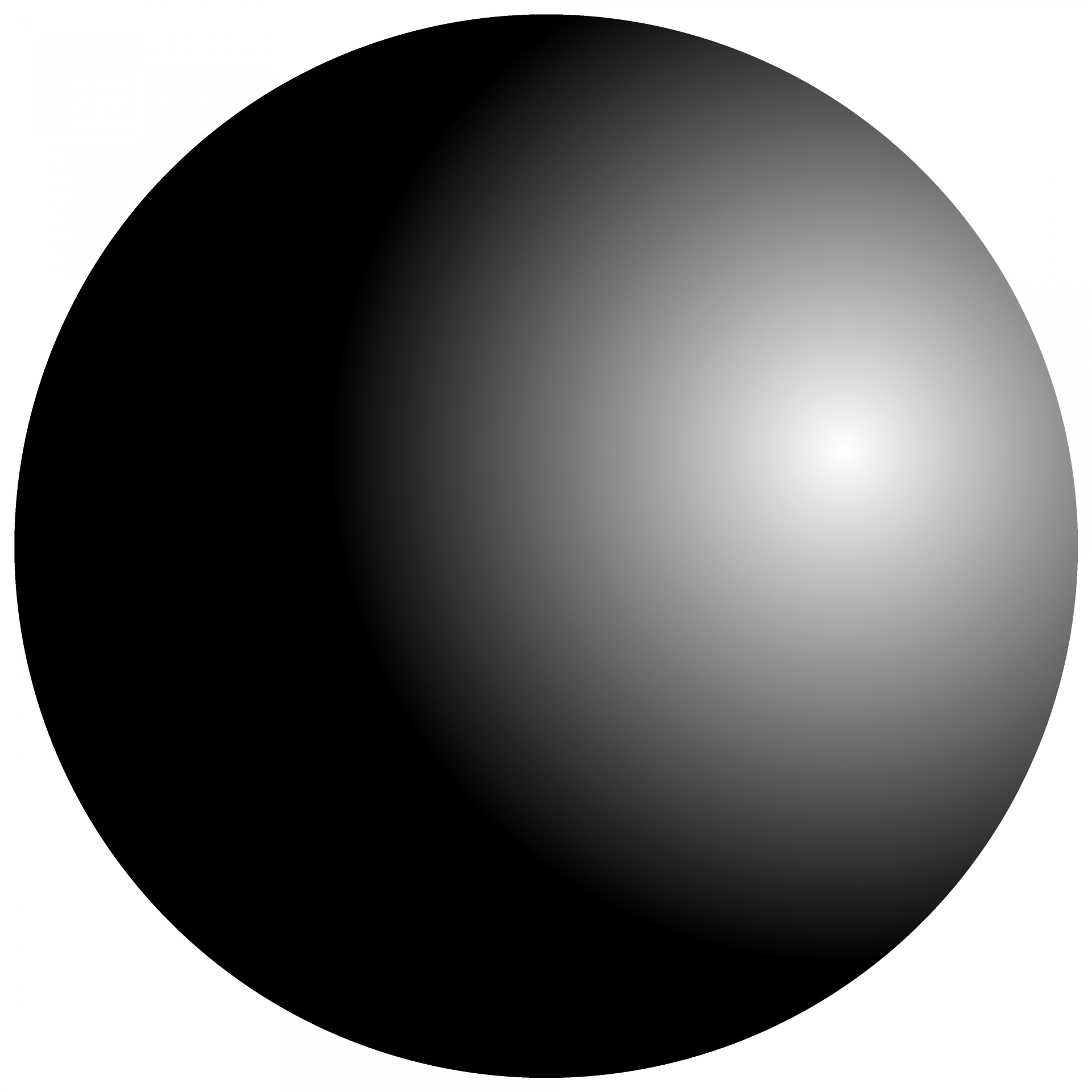} & Sphere 1 &  \begin{tabular}{@{}c@{}} $r$=199 nm \\ $\epsilon_r=2.19^2$ \end{tabular}  &%
\begin{tabular}{c}
\vspace*{-0.3cm}
\tabularnewline
1.99e-03 \tabularnewline
\vspace*{-0.3cm}
\tabularnewline
\end{tabular} &  2.76  &                     \begin{tabular}{@{}c@{}} 0.4578-0.4982i \\ 0.4548-0.4980i \end{tabular}  & \begin{tabular}{@{}c@{}} 0.0037-0.0610i \\ 0.0002-0.0143i \end{tabular} & \begin{tabular}{@{}c@{}} 0.0000-0.0023i  \\ 0.0000-0.0003i \end{tabular}\tabularnewline
\hline
 \includegraphics[trim={1cm 1cm 1cm -1cm},clip,scale=0.12]{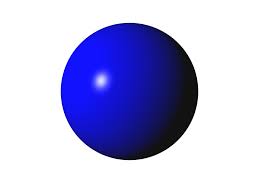} & Sphere 2 &  \begin{tabular}{@{}c@{}} $r$=180 nm \\ $\epsilon_r=2.42^2$ \end{tabular}  & %
\begin{tabular}{c}
\vspace*{-0.3cm}
\tabularnewline
1.04e-03\tabularnewline
\vspace*{-0.3cm}
\tabularnewline
\end{tabular} &  2.41 &                     \begin{tabular}{@{}c@{}} 0.3979-0.4895i \\ 0.4009-0.4901i \end{tabular}  & \begin{tabular}{@{}c@{}}  0.0017-0.0410i \\ 0.0001-0.0095i \end{tabular} & \begin{tabular}{@{}c@{}} 0.0000-0.0013i \\ 0.0000-0.0002i \end{tabular}\tabularnewline
\hline
 \includegraphics[trim={0.01cm 1cm 0.01cm -2cm},clip,scale=0.08]{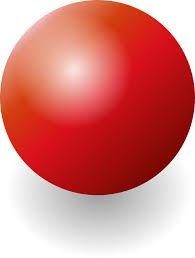} & Sphere 3 &  \begin{tabular}{@{}c@{}} $r$=161 nm \\ $\epsilon_r=2.71^2$ \end{tabular}  &%
\begin{tabular}{c}
\vspace*{-0.3cm}
\tabularnewline
5.64e-04\tabularnewline
\vspace*{-0.3cm}
\tabularnewline
\end{tabular} & 1.84 &                     \begin{tabular}{@{}c@{}} 0.3035-0.4598i \\ 0.3094-0.4623i \end{tabular}  & \begin{tabular}{@{}c@{}} 0.0007-0.0256i  \\ 0.0000-0.0059i \end{tabular} & \begin{tabular}{@{}c@{}} 0.0000-0.0006i \\ 0.0000-0.0001i \end{tabular} \tabularnewline
\hline
 \includegraphics[trim={1cm 1cm 1cm -1cm},clip,scale=0.1]{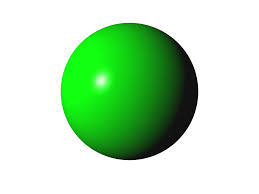} & Sphere 4 & \begin{tabular}{@{}c@{}} $r$=142 nm \\ $\epsilon_r=3.08^2$ \end{tabular}  & %
\begin{tabular}{c}
\vspace*{-0.3cm}
\tabularnewline
2.94e-04 \tabularnewline
\vspace*{-0.3cm}
\tabularnewline
\end{tabular} &  1.17 &                     \begin{tabular}{@{}c@{}}  0.1927-0.3945i \\ 0.1955-0.3966i  \end{tabular}  & \begin{tabular}{@{}c@{}}  0.0002-0.0147i \\ 0.0000-0.0033i \end{tabular} & \begin{tabular}{@{}c@{}}  0.0000-0.0003i\\0.0000-0.0000i \end{tabular}\tabularnewline
\hline
 \includegraphics[trim={1cm 0.2cm 0.5cm -5cm},clip,scale=0.03]{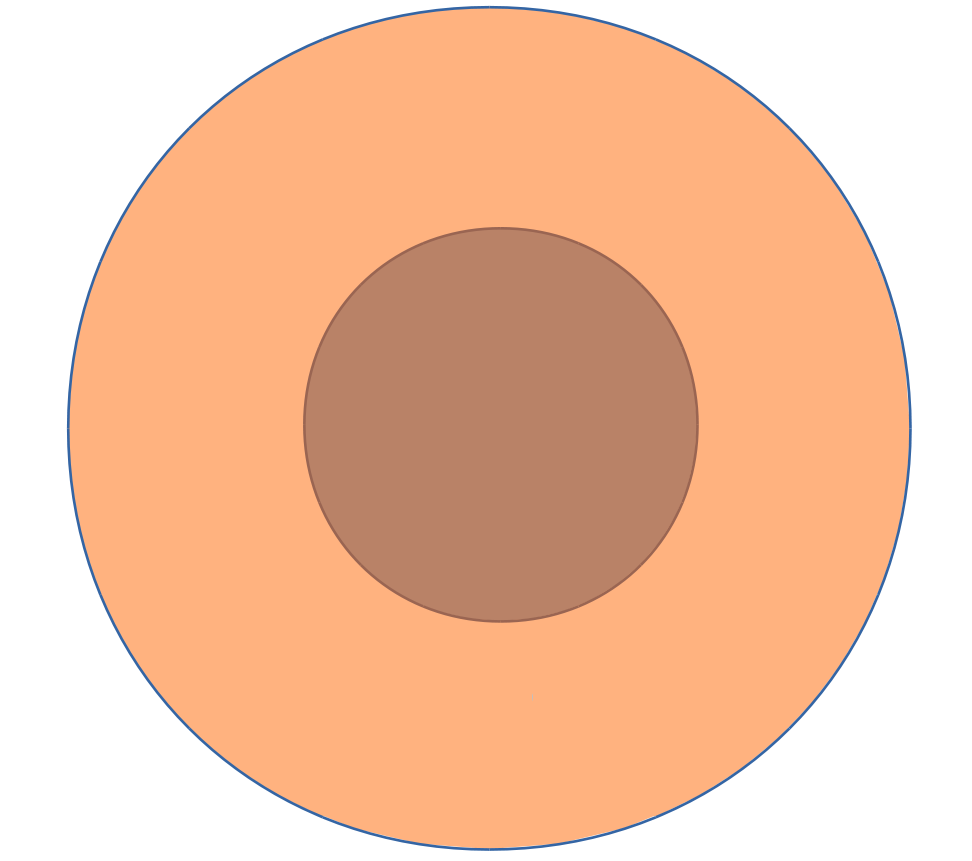} & Core-Shell &  \begin{tabular}{@{}c@{}} $r_{\text{inner}}$=105 nm, $r_{\text{outer}}$=225 nm, \\ $\epsilon_{\text{inner}}=4^2$, $\epsilon_{\text{outer}}=2.7^2$  \end{tabular}  &%
\begin{tabular}{c}
\vspace*{-0.3cm}
\tabularnewline
4.57e-05 \tabularnewline
\vspace*{-0.3cm}
\tabularnewline
\end{tabular} & %
\begin{tabular}{c}
\vspace*{-0.3cm}
\tabularnewline
1.55 \tabularnewline
\vspace*{-0.3cm}
\tabularnewline
\end{tabular} &                         \begin{tabular}{@{}c@{}} 0.2118+0.4086i \\ 0.2117+0.4085i\end{tabular}  & \begin{tabular}{@{}c@{}} 0.0277-0.1641i \\ 0.0278-0.1643i\end{tabular} & \begin{tabular}{@{}c@{}} 0.0000-0.0065i \\ 0.0000-0.0020i \end{tabular} \tabularnewline
\hline
\end{tabular}
\par\end{centering}
\end{table*}
The electromagnetic duality symmetry of electric and magnetic fields in the free space Maxwell's equations is typically broken in light-matter interactions. In order to quantify the breaking of the symmetry by the material object, the T-matrix is changed to the helicity basis:
\begin{eqnarray}
\mathbf{T}^{\mathrm{helicity}}	& = &	\left(\begin{array}{cc}
\mathbf{T}_{\mathrm{++}}^{\mathrm{helicity}} & \mathbf{T}_{\mathrm{+-}}^{\mathrm{helicity}}\\
\mathbf{T}_{\mathrm{-+}}^{\mathrm{helicity}} & \mathbf{T}_{\mathrm{--}}^{\mathrm{helicity}}
\end{array}\right) \\ & = & \frac{1}{2}\left(\begin{array}{cc}
\mathbf{1} & \mathbf{1}\\
\mathbf{1} & \mathbf{-1}
\end{array}\right) \mathbf{T}\left(\begin{array}{cc}
\mathbf{1} & \mathbf{1}\\
\mathbf{1} & \mathbf{-1}
\end{array}\right),  \nonumber
 \end{eqnarray}
where $\mathbf{1}$ is the identity matrix of the size of the $\mathbf{T}_{ab}$ in Eq. (\ref{eq:T}). Then, the duality breaking $\cancel{D}$ can be defined as the ratio of the sum of the squares of all helicity changing matrix elements divided by the sum of the squares of all the matrix elements (\cite[Eq. 2]{Fernandez-Corbaton2015}):
{\small
\begin{eqnarray}
	\label{eq:ddash}
	\cancel{D}	=\frac{\Tr{\left({\mathbf{T}_{\mathrm{+-}}^{\mathrm{helicity}}}\right)^\dagger{\mathbf{T}_{\mathrm{+-}}^{\mathrm{helicity}}}+\left({\mathbf{T}_{\mathrm{-+}}^{\mathrm{helicity}}}\right)^\dagger{\mathbf{T}_{\mathrm{-+}}^{\mathrm{helicity}}}}}{\Tr{\left({\mathbf{T}_{}^{\mathrm{helicity}}}\right)^\dagger{\mathbf{T}_{}^{\mathrm{helicity}}}}}\label{db},
 \end{eqnarray}
 }
where $\Tr{A}$ denotes the trace of $A$. The duality breaking ($\cancel{D}$) is a number between 0 and 1, and is 0 for a perfectly dual object. For an isotropic particle $\cancel{D}$ can be expressed in terms of the Mie coefficients and reads as:
\begin{eqnarray}
	\cancel{D}	=	\frac{1}{2}\frac{\sum_{j=1}^{\infty}\left|a_{j}-b_{j}\right|^2(2j+1)}{\sum_{j=1}^{\infty}\left(\left|b_{j}\right|^{2}+\left|a_{j}\right|^{2}\right)(2j+1)},
 \end{eqnarray}
\noindent where $a_j$ and $b_j$ are the electric and magnetic Mie coefficients. A perfectly dual sphere has $a_j=b_j$ for all $j$.

Another quantity of importance used to characterize the scattering properties of an object is the total interaction strength defined as
\begin{eqnarray}
	C=\Tr{\left({\mathbf{T}_{}^{\mathrm{helicity}}}\right)^\dagger{\mathbf{T}_{}^{\mathrm{helicity}}}},
\end{eqnarray}
which appears in the denominator of Eq. (\ref{eq:ddash}). Unlike the scattering cross section, the total interaction strength is an illumination independent measure of the strength of the interaction of an object with the electromagnetic field. For an isotropic particle, the total interaction strength, corresponds to the normalized [to $\lambda^2/(2\pi)$] scattering cross section \cite{rahimzadegan2017Fundamental,ruan2010superscattering}. 

\begin{figure}
\begin{centering}
	\includegraphics[width=\linewidth]{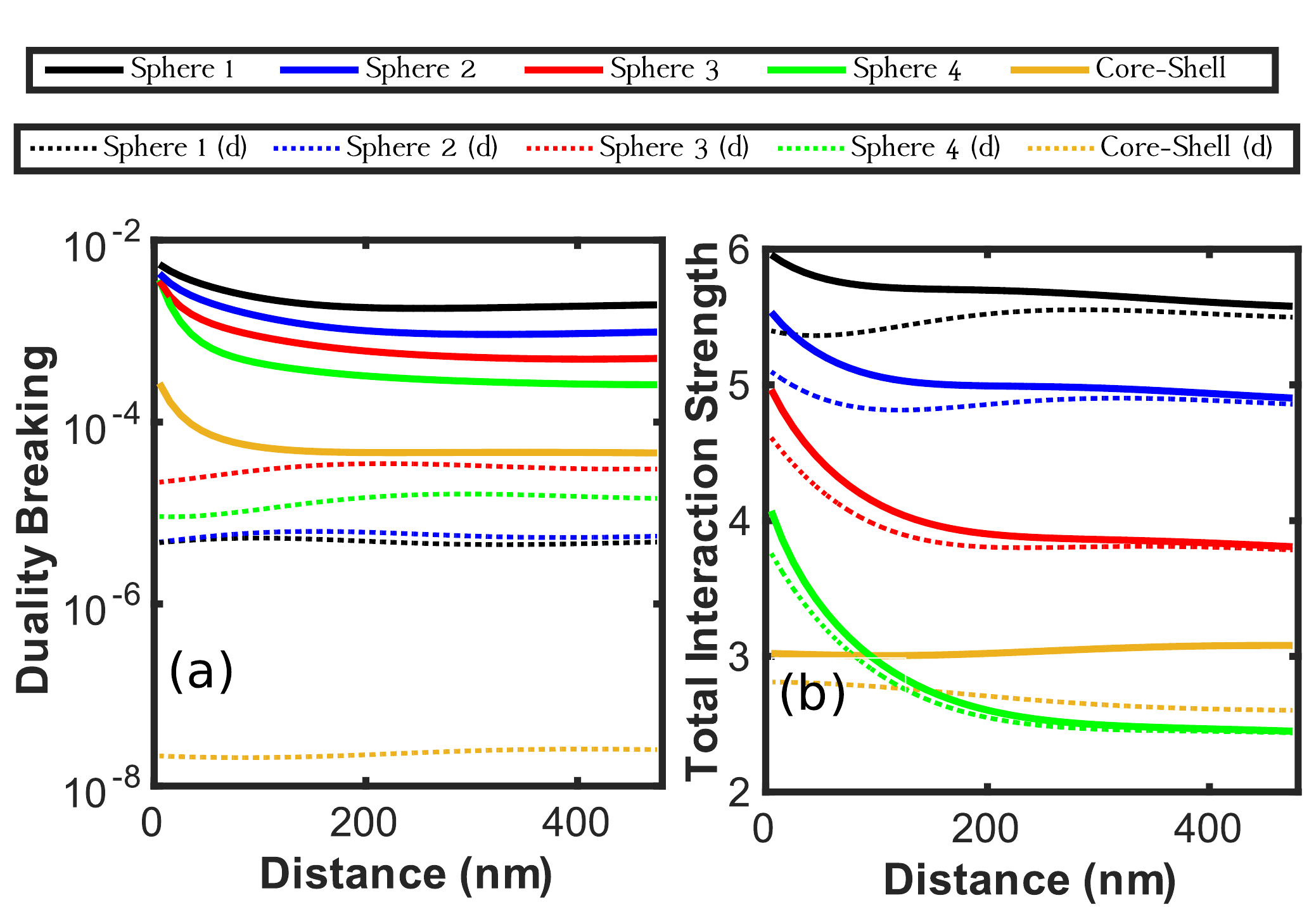}
\par\end{centering}
\protect\caption{Duality breaking (a) and total interaction strength (b) of dimers made out of the five particles in Table \ref{table:Force} as a function of the inter-particle distance. The solid lines show the results when all the multipolar orders of the response of the spheres are considered. The dotted lines show the results when keeping only their dipolar response.\label{fig: distance} }
\end{figure}

The use of individual particles as building blocks for larger structures and effective bulk media is one of the main ideas in metamaterials. In the context of this Letter, we are interested in the duality properties of composite objects. If each individual particle in an ensemble is perfectly dual, we are guaranteed that the ensemble will be perfectly dual as well. But when approximately dual particles are placed close to each other, it is not immediately clear what is the duality breaking of the composite object. In order to address this question, as the first step, we have computed $\cancel{D}$ as a function of the inter-particle distance for four dimers made with approximately dual homogeneous dielectric spheres. The four spheres are similar\footnote{They are conformally rescaled versions of the spheres used in Ref. \onlinecite{Fernandez-Corbaton2015}. The rescaling of radii and permittivities is done in order to change the surrounding medium from one with permittivity $\epsilon=(1.3)^2$ to vacuum. The duality breaking $\cancel{D}$ of the spheres is invariant under the rescaling.} to those used in Ref. \onlinecite{Fernandez-Corbaton2015}. The operating wavelength is $\lambda=1\mu\text{m}$. The radii, relative permittivities, duality breaking $\cancel{D}$, total interaction strengths $C$, and the first three pairs of electric and magnetic Mie coefficients for the four homogeneous spheres are shown in Table \ref{table:Force}. The solid lines labeled as Sphere 1 to 4 in Figs.~\ref{fig: distance}(a) and \ref{fig: distance}(b) show $\cancel{D}$ and $C$ as a function of the inter-particle distance, respectively. We observe that the duality symmetry of the dimers degrades very rapidly at short distances. The Mie coefficients for the individual spheres in Table \ref{table:Force} give us the hint for understanding this behavior. The electric and magnetic dipolar coefficients of each sphere are quite similar to each other, but the quadrupolar and octopolar ones are not. This, together with the known enhancement of the relative contribution of higher order multipoles in near field interactions \cite{Tojo2004,Tojo2005,FerCor2016b}, explains the results. We confirm this hypothesis by artificially truncating the responses of the four spheres to just the dipolar terms and repeating the calculations of the $\cancel{D}$ of the dimers. The dotted lines labeled Sphere 1(d) to 4(d) show the results in Fig.~\ref{fig: distance}. The rapid increase in $\cancel{D}$ as the distance decreases disappears for pure dipolar interactions. Please note that we have used a global multipolar expansion order of $N=8$, which is accurate and converging for all the numerical calculations in the paper.
The fact that the parameters of the four spheres were already chosen in Ref. \onlinecite{Fernandez-Corbaton2015} to maximize their duality indicates that a different kind of particle is needed in order to improve the situation seen in  Fig.~\ref{fig: distance}(a).  We now explore the benefits of the extra degrees of freedom in core-shell particles. Core-shell spherical particles are being experimentally investigated as the building blocks of larger objects \cite{Malassis2013,Ponsinet2014}. We have used the Particle Swarm Optimization (PSO) method \cite{lee2008modern} to design a core-shell particle that has a minimum $\cancel{D}$. In order to provide some initial starting points to the optimization process, a coarse sweeping simulation was previously ran. The final design parameters and properties of the proposed particle are shown in the last row of Table \ref{table:Force}. We note that for the quadrupolar and octopolar orders, the electric Mie coefficients are much more similar to the magnetic ones than in the case of the homogeneous spheres. This allows the corresponding dimer, labeled Core-Shell in Fig.~\ref{fig: distance}, to keep its $\cancel{D}$ more than one order of magnitude below that of other four dimers for all (including vanishing) inter-particle distances. With respect to $C$, an interesting observation in Fig.~\ref{fig: distance}(b) is the relative flatness of the total interaction strength as a function of the distance. This should be studied in the future. It is worth mentioning that the PSO was targeted at minimizing $\cancel{D}$, and that more complex optimization targets could include both $\cancel{D}$ and $C$ simultaneously. 

\begin{figure}
\begin{centering}
\includegraphics[width=\linewidth]{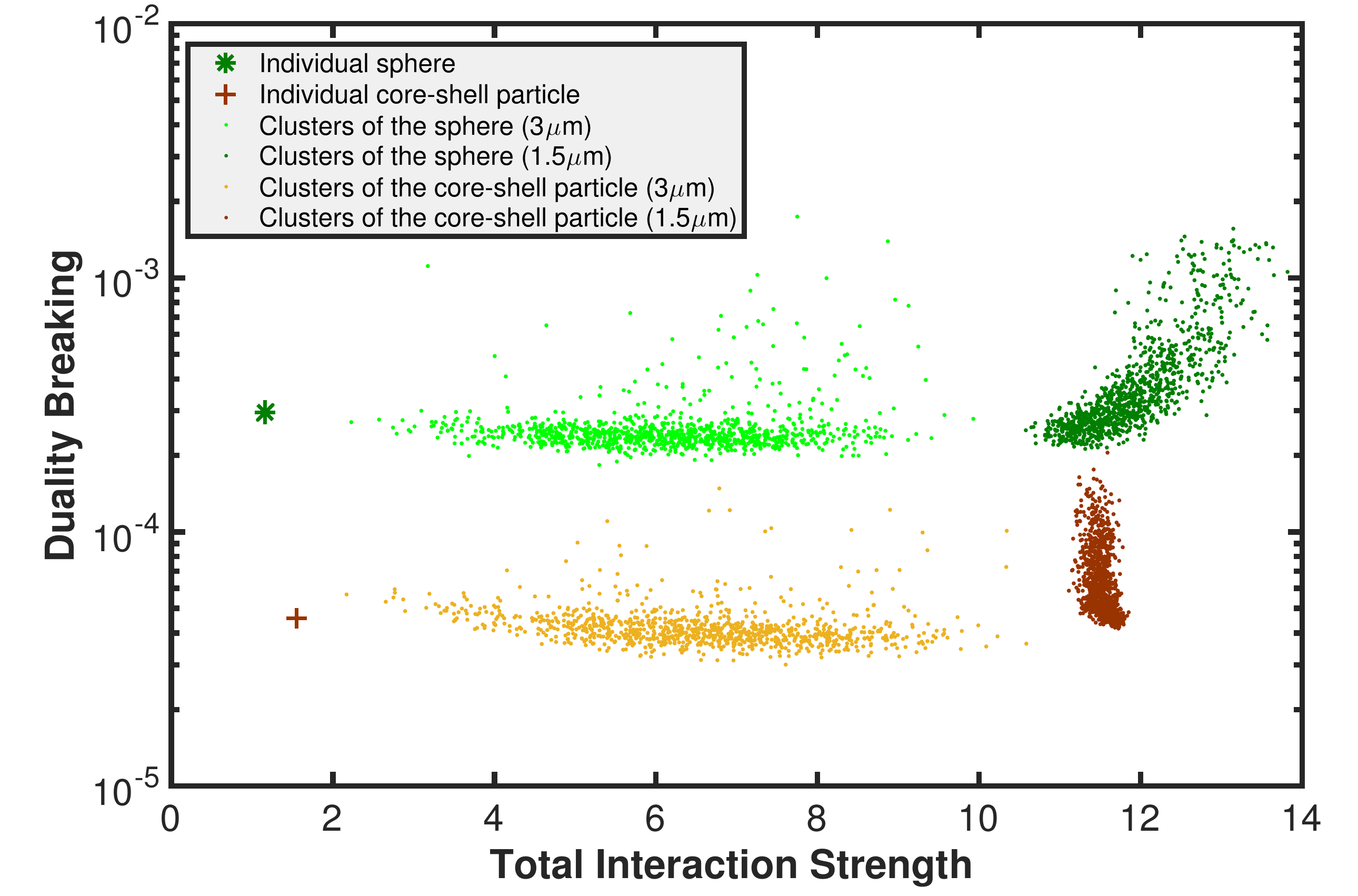}
\par\end{centering}
\protect\caption{Scatter plots of duality breaking versus total interaction strength for 1000 realizations of randomly arranged clusters of eight core-shells and eight Sphere 4s. Two different confining cubic volumes of side 1.5 and 3$\mu$m have been used. The green star, light green dots, and green dots correspond to the individual Sphere 4, random arrangements of eight of them in a cube of 3$\mu$m side, and random arrangements of eight of them in a cube of 1.5$\mu$m side, respectively. The brown cross, light brown dots, and brown dots correspond to the core-shell particle, random arrangements of eight of them in a cube of 3$\mu$m side, and random arrangements of eight of them in a cube of 1.5$\mu$m side, respectively.\label{fig: cluster} }
\end{figure}

One clear application of approximately dual particles is the realization of approximately dual bulk materials. In this application, several particles will be close to each other. In order to investigate the duality properties of an ensemble of several particles, we randomly arrange eight particles of the same type confined to a cubic volume and compute $\cancel{D}$ and $C$ for the ensemble. We use the designed core-shell particle, and for comparison, the best of the four spheres in terms of duality symmetry (Sphere 4). We simulate 1000 independent random ensembles for each of the two confining cubic volumes of sides 1.5 and 3$\mu$m. The results are shown in Fig.~\ref{fig: cluster}. The order of magnitude improvement in $\cancel{D}$ for the individual particles is kept in the ensembles for both volume sizes. We also observe that the lower bound of duality breaking is largely independent of the total interaction strength for both kinds of particles, even for different confining volumes. The results seen at the two right-most point clouds correspond to the tighter packed cases, and show that the total interaction strength varies much less for core-shell particles than homogeneous ones. This behavior is consistent with the finding from Fig. \ref{fig: distance}(b) in that the total interaction strength of dimers made out of core-shell particles is largely independent of the distance between the particles, even for very short distances at which the total interaction strengths of its homogeneous counterparts increase rapidly.

In conclusion, we have shown that the extra degrees of freedom of dielectric core-shell spheres can be used to design isotropic particles whose duality symmetry is an order of magnitude better than for dielectric homogeneous spheres. Importantly, the same improvement is maintained in the overall duality of composite objects when the core-shell particles are used as building blocks instead of the homogeneous spheres. It is reasonable to expect further improvements for each extra spherical shell, and that the use of shells helps the design of core-shell helices \cite{Kosters2017} of high duality. The results of this work are a step towards the realization of composite objects and effective bulk materials with a high degree of duality symmetry, which have technologically important applications like artificial optical activity, zero-backscattering, metamaterials for transformation optics, wave-front control, and topologically protected light propagation.

\section*{Acknowledgements}
This work has been partly funded by the German Science Foundation within the priority program SPP1839 Tailored Disorder (RO 3640/7-1).

%


\begin{thebibliography}{32}%
\makeatletter
\providecommand \@ifxundefined [1]{%
 \@ifx{#1\undefined}
}%
\providecommand \@ifnum [1]{%
 \ifnum #1\expandafter \@firstoftwo
 \else \expandafter \@secondoftwo
 \fi
}%
\providecommand \@ifx [1]{%
 \ifx #1\expandafter \@firstoftwo
 \else \expandafter \@secondoftwo
 \fi
}%
\providecommand \natexlab [1]{#1}%
\providecommand \enquote  [1]{``#1''}%
\providecommand \bibnamefont  [1]{#1}%
\providecommand \bibfnamefont [1]{#1}%
\providecommand \citenamefont [1]{#1}%
\providecommand \href@noop [0]{\@secondoftwo}%
\providecommand \href [0]{\begingroup \@sanitize@url \@href}%
\providecommand \@href[1]{\@@startlink{#1}\@@href}%
\providecommand \@@href[1]{\endgroup#1\@@endlink}%
\providecommand \@sanitize@url [0]{\catcode `\\12\catcode `\$12\catcode
  `\&12\catcode `\#12\catcode `\^12\catcode `\_12\catcode `\%12\relax}%
\providecommand \@@startlink[1]{}%
\providecommand \@@endlink[0]{}%
\providecommand \url  [0]{\begingroup\@sanitize@url \@url }%
\providecommand \@url [1]{\endgroup\@href {#1}{\urlprefix }}%
\providecommand \urlprefix  [0]{URL }%
\providecommand \Eprint [0]{\href }%
\providecommand \doibase [0]{http://dx.doi.org/}%
\providecommand \selectlanguage [0]{\@gobble}%
\providecommand \bibinfo  [0]{\@secondoftwo}%
\providecommand \bibfield  [0]{\@secondoftwo}%
\providecommand \translation [1]{[#1]}%
\providecommand \BibitemOpen [0]{}%
\providecommand \bibitemStop [0]{}%
\providecommand \bibitemNoStop [0]{.\EOS\space}%
\providecommand \EOS [0]{\spacefactor3000\relax}%
\providecommand \BibitemShut  [1]{\csname bibitem#1\endcsname}%
\let\auto@bib@innerbib\@empty
\bibitem [{\citenamefont {Wagner}\ and\ \citenamefont
  {Lynch}(1963)}]{Wagner1963}%
  \BibitemOpen
  \bibfield  {author} {\bibinfo {author} {\bibfnamefont {R.~J.}\ \bibnamefont
  {Wagner}}\ and\ \bibinfo {author} {\bibfnamefont {P.~J.}\ \bibnamefont
  {Lynch}},\ }\href {\doibase 10.1103/PhysRev.131.21} {\bibfield  {journal}
  {\bibinfo  {journal} {Phys. Rev.}\ }\textbf {\bibinfo {volume} {131}},\
  \bibinfo {pages} {21} (\bibinfo {year} {1963})}\BibitemShut {NoStop}%
\bibitem [{\citenamefont {Kerker}\ \emph {et~al.}(1983)\citenamefont {Kerker},
  \citenamefont {Wang},\ and\ \citenamefont {Giles}}]{Kerker1983}%
  \BibitemOpen
  \bibfield  {author} {\bibinfo {author} {\bibfnamefont {M.}~\bibnamefont
  {Kerker}}, \bibinfo {author} {\bibfnamefont {D.~S.}\ \bibnamefont {Wang}}, \
  and\ \bibinfo {author} {\bibfnamefont {C.~L.}\ \bibnamefont {Giles}},\
  }\href@noop {} {\bibfield  {journal} {\bibinfo  {journal} {J. Opt. Soc. Am.}\
  }\textbf {\bibinfo {volume} {73}},\ \bibinfo {pages} {765} (\bibinfo {year}
  {1983})}\BibitemShut {NoStop}%
\bibitem [{\citenamefont {Karilainen}\ and\ \citenamefont
  {Tretyakov}(2012)}]{Karilainen2012}%
  \BibitemOpen
  \bibfield  {author} {\bibinfo {author} {\bibfnamefont {A.~O.}\ \bibnamefont
  {Karilainen}}\ and\ \bibinfo {author} {\bibfnamefont {S.~A.}\ \bibnamefont
  {Tretyakov}},\ }\href {\doibase 10.1109/TAP.2012.2207069} {\bibfield
  {journal} {\bibinfo  {journal} {IEEE Trans. Antennas Propagat.}\ }\textbf
  {\bibinfo {volume} {60}},\ \bibinfo {pages} {4449} (\bibinfo {year}
  {2012})}\BibitemShut {NoStop}%
\bibitem [{\citenamefont {Zambrana-Puyalto}\ \emph
  {et~al.}(2013{\natexlab{a}})\citenamefont {Zambrana-Puyalto}, \citenamefont
  {Fernandez-Corbaton}, \citenamefont {Juan}, \citenamefont {Vidal},\ and\
  \citenamefont {Molina-Terriza}}]{Zambrana2013}%
  \BibitemOpen
  \bibfield  {author} {\bibinfo {author} {\bibfnamefont {X.}~\bibnamefont
  {Zambrana-Puyalto}}, \bibinfo {author} {\bibfnamefont {I.}~\bibnamefont
  {Fernandez-Corbaton}}, \bibinfo {author} {\bibfnamefont {M.~L.}\ \bibnamefont
  {Juan}}, \bibinfo {author} {\bibfnamefont {X.}~\bibnamefont {Vidal}}, \ and\
  \bibinfo {author} {\bibfnamefont {G.}~\bibnamefont {Molina-Terriza}},\ }\href
  {\doibase 10.1364/OL.38.001857} {\bibfield  {journal} {\bibinfo  {journal}
  {Opt. Lett.}\ }\textbf {\bibinfo {volume} {38}},\ \bibinfo {pages} {1857}
  (\bibinfo {year} {2013}{\natexlab{a}})}\BibitemShut {NoStop}%
\bibitem [{\citenamefont {Fernandez-Corbaton}(2013)}]{FerCor2013c}%
  \BibitemOpen
  \bibfield  {author} {\bibinfo {author} {\bibfnamefont {I.}~\bibnamefont
  {Fernandez-Corbaton}},\ }\href {\doibase 10.1364/OE.21.029885} {\bibfield
  {journal} {\bibinfo  {journal} {Opt. Express}\ }\textbf {\bibinfo {volume}
  {21}},\ \bibinfo {pages} {29885} (\bibinfo {year} {2013})}\BibitemShut
  {NoStop}%
\bibitem [{\citenamefont {Zhang}\ \emph {et~al.}(2015)\citenamefont {Zhang},
  \citenamefont {Nieto-Vesperinas},\ and\ \citenamefont {Sáenz}}]{Zhang2017}%
  \BibitemOpen
  \bibfield  {author} {\bibinfo {author} {\bibfnamefont {Y.}~\bibnamefont
  {Zhang}}, \bibinfo {author} {\bibfnamefont {M.}~\bibnamefont
  {Nieto-Vesperinas}}, \ and\ \bibinfo {author} {\bibfnamefont {J.~J.}\
  \bibnamefont {Sáenz}},\ }\href@noop {} {\bibfield  {journal} {\bibinfo
  {journal} {Journal of Optics}\ }\textbf {\bibinfo {volume} {17}},\ \bibinfo
  {pages} {105612} (\bibinfo {year} {2015})}\BibitemShut {NoStop}%
\bibitem [{\citenamefont {Fernandez-Corbaton}\ \emph
  {et~al.}(2015)\citenamefont {Fernandez-Corbaton}, \citenamefont {Fruhnert},\
  and\ \citenamefont {Rockstuhl}}]{Fernandez-Corbaton2015}%
  \BibitemOpen
  \bibfield  {author} {\bibinfo {author} {\bibfnamefont {I.}~\bibnamefont
  {Fernandez-Corbaton}}, \bibinfo {author} {\bibfnamefont {M.}~\bibnamefont
  {Fruhnert}}, \ and\ \bibinfo {author} {\bibfnamefont {C.}~\bibnamefont
  {Rockstuhl}},\ }\href@noop {} {\bibfield  {journal} {\bibinfo  {journal} {ACS
  Photonics}\ } (\bibinfo {year} {2015})}\BibitemShut {NoStop}%
\bibitem [{\citenamefont {Leonhardt}\ and\ \citenamefont
  {Philbin}(2009)}]{Leonhardt2009}%
  \BibitemOpen
  \bibfield  {author} {\bibinfo {author} {\bibfnamefont {U.}~\bibnamefont
  {Leonhardt}}\ and\ \bibinfo {author} {\bibfnamefont {T.~G.}\ \bibnamefont
  {Philbin}},\ }\href
  {http://www.sciencedirect.com/science/article/pii/S0079663808002023}
  {\bibfield  {journal} {\bibinfo  {journal} {Progress in Optics}\ }\textbf
  {\bibinfo {volume} {53}},\ \bibinfo {pages} {69} (\bibinfo {year}
  {2009})}\BibitemShut {NoStop}%
\bibitem [{\citenamefont {Thompson}\ and\ \citenamefont
  {Frauendiener}(2010)}]{Thompson2010}%
  \BibitemOpen
  \bibfield  {author} {\bibinfo {author} {\bibfnamefont {R.~T.}\ \bibnamefont
  {Thompson}}\ and\ \bibinfo {author} {\bibfnamefont {J.}~\bibnamefont
  {Frauendiener}},\ }\href {\doibase 10.1103/PhysRevD.82.124021} {\bibfield
  {journal} {\bibinfo  {journal} {Phys. Rev. D}\ }\textbf {\bibinfo {volume}
  {82}},\ \bibinfo {pages} {124021} (\bibinfo {year} {2010})}\BibitemShut
  {NoStop}%
\bibitem [{\citenamefont {Thompson}\ \emph {et~al.}(2011)\citenamefont
  {Thompson}, \citenamefont {Cummer},\ and\ \citenamefont
  {Frauendiener}}]{Thompson2011}%
  \BibitemOpen
  \bibfield  {author} {\bibinfo {author} {\bibfnamefont {R.~T.}\ \bibnamefont
  {Thompson}}, \bibinfo {author} {\bibfnamefont {S.~A.}\ \bibnamefont
  {Cummer}}, \ and\ \bibinfo {author} {\bibfnamefont {J.}~\bibnamefont
  {Frauendiener}},\ }\href {http://stacks.iop.org/2040-8986/13/i=2/a=024008}
  {\bibfield  {journal} {\bibinfo  {journal} {Journal of Optics}\ }\textbf
  {\bibinfo {volume} {13}},\ \bibinfo {pages} {024008} (\bibinfo {year}
  {2011})}\BibitemShut {NoStop}%
\bibitem [{\citenamefont {Fernandez-Corbaton}\ and\ \citenamefont
  {Molina-Terriza}(2013)}]{FerCor2013}%
  \BibitemOpen
  \bibfield  {author} {\bibinfo {author} {\bibfnamefont {I.}~\bibnamefont
  {Fernandez-Corbaton}}\ and\ \bibinfo {author} {\bibfnamefont
  {G.}~\bibnamefont {Molina-Terriza}},\ }\href {\doibase
  10.1103/PhysRevB.88.085111} {\bibfield  {journal} {\bibinfo  {journal} {Phys.
  Rev. B}\ }\textbf {\bibinfo {volume} {88}},\ \bibinfo {pages} {085111}
  (\bibinfo {year} {2013})}\BibitemShut {NoStop}%
\bibitem [{\citenamefont {Pfeiffer}\ and\ \citenamefont
  {Grbic}(2013)}]{Pfeiffer2013}%
  \BibitemOpen
  \bibfield  {author} {\bibinfo {author} {\bibfnamefont {C.}~\bibnamefont
  {Pfeiffer}}\ and\ \bibinfo {author} {\bibfnamefont {A.}~\bibnamefont
  {Grbic}},\ }\href {\doibase 10.1103/PhysRevLett.110.197401} {\bibfield
  {journal} {\bibinfo  {journal} {Phys. Rev. Lett.}\ }\textbf {\bibinfo
  {volume} {110}},\ \bibinfo {pages} {197401} (\bibinfo {year}
  {2013})}\BibitemShut {NoStop}%
\bibitem [{\citenamefont {Chong}\ \emph {et~al.}(2016)\citenamefont {Chong},
  \citenamefont {Wang}, \citenamefont {Staude}, \citenamefont {James},
  \citenamefont {Dominguez}, \citenamefont {Liu}, \citenamefont {Subramania},
  \citenamefont {Decker}, \citenamefont {Neshev}, \citenamefont {Brener},\ and\
  \citenamefont {Kivshar}}]{Chong2016}%
  \BibitemOpen
  \bibfield  {author} {\bibinfo {author} {\bibfnamefont {K.~E.}\ \bibnamefont
  {Chong}}, \bibinfo {author} {\bibfnamefont {L.}~\bibnamefont {Wang}},
  \bibinfo {author} {\bibfnamefont {I.}~\bibnamefont {Staude}}, \bibinfo
  {author} {\bibfnamefont {A.~R.}\ \bibnamefont {James}}, \bibinfo {author}
  {\bibfnamefont {J.}~\bibnamefont {Dominguez}}, \bibinfo {author}
  {\bibfnamefont {S.}~\bibnamefont {Liu}}, \bibinfo {author} {\bibfnamefont
  {G.~S.}\ \bibnamefont {Subramania}}, \bibinfo {author} {\bibfnamefont
  {M.}~\bibnamefont {Decker}}, \bibinfo {author} {\bibfnamefont {D.~N.}\
  \bibnamefont {Neshev}}, \bibinfo {author} {\bibfnamefont {I.}~\bibnamefont
  {Brener}}, \ and\ \bibinfo {author} {\bibfnamefont {Y.~S.}\ \bibnamefont
  {Kivshar}},\ }\href {\doibase 10.1021/acsphotonics.5b00678} {\bibfield
  {journal} {\bibinfo  {journal} {ACS Photonics}\ }\textbf {\bibinfo {volume}
  {3}},\ \bibinfo {pages} {514} (\bibinfo {year} {2016})}\BibitemShut {NoStop}%
\bibitem [{\citenamefont {Ziolkowski}(2017)}]{Ziolkowski2017}%
  \BibitemOpen
  \bibfield  {author} {\bibinfo {author} {\bibfnamefont {R.~W.}\ \bibnamefont
  {Ziolkowski}},\ }\href {\doibase 10.1103/PhysRevX.7.031017} {\bibfield
  {journal} {\bibinfo  {journal} {Phys. Rev. X}\ }\textbf {\bibinfo {volume}
  {7}},\ \bibinfo {pages} {031017} (\bibinfo {year} {2017})}\BibitemShut
  {NoStop}%
\bibitem [{\citenamefont {Fernandez-Corbaton}\ \emph
  {et~al.}(2016{\natexlab{a}})\citenamefont {Fernandez-Corbaton}, \citenamefont
  {Fruhnert},\ and\ \citenamefont {Rockstuhl}}]{FerCor2016}%
  \BibitemOpen
  \bibfield  {author} {\bibinfo {author} {\bibfnamefont {I.}~\bibnamefont
  {Fernandez-Corbaton}}, \bibinfo {author} {\bibfnamefont {M.}~\bibnamefont
  {Fruhnert}}, \ and\ \bibinfo {author} {\bibfnamefont {C.}~\bibnamefont
  {Rockstuhl}},\ }\href {\doibase 10.1103/PhysRevX.6.031013} {\bibfield
  {journal} {\bibinfo  {journal} {Phys. Rev. X}\ }\textbf {\bibinfo {volume}
  {6}},\ \bibinfo {pages} {031013} (\bibinfo {year}
  {2016}{\natexlab{a}})}\BibitemShut {NoStop}%
\bibitem [{\citenamefont {Rahimzadegan}\ \emph {et~al.}(2016)\citenamefont
  {Rahimzadegan}, \citenamefont {Fruhnert}, \citenamefont {Alaee},
  \citenamefont {Fernandez-Corbaton},\ and\ \citenamefont
  {Rockstuhl}}]{rahimzadegan2016optical}%
  \BibitemOpen
  \bibfield  {author} {\bibinfo {author} {\bibfnamefont {A.}~\bibnamefont
  {Rahimzadegan}}, \bibinfo {author} {\bibfnamefont {M.}~\bibnamefont
  {Fruhnert}}, \bibinfo {author} {\bibfnamefont {R.}~\bibnamefont {Alaee}},
  \bibinfo {author} {\bibfnamefont {I.}~\bibnamefont {Fernandez-Corbaton}}, \
  and\ \bibinfo {author} {\bibfnamefont {C.}~\bibnamefont {Rockstuhl}},\ }\href
  {\doibase 10.1103/PhysRevB.94.125123} {\bibfield  {journal} {\bibinfo
  {journal} {Phys. Rev. B}\ }\textbf {\bibinfo {volume} {94}},\ \bibinfo
  {pages} {125123} (\bibinfo {year} {2016})}\BibitemShut {NoStop}%
\bibitem [{\citenamefont {Schmidt}\ \emph {et~al.}(2015)\citenamefont
  {Schmidt}, \citenamefont {Aizpurua}, \citenamefont {Zambrana-Puyalto},
  \citenamefont {Vidal}, \citenamefont {Molina-Terriza},\ and\ \citenamefont
  {S\'aenz}}]{Schmidt2015}%
  \BibitemOpen
  \bibfield  {author} {\bibinfo {author} {\bibfnamefont {M.~K.}\ \bibnamefont
  {Schmidt}}, \bibinfo {author} {\bibfnamefont {J.}~\bibnamefont {Aizpurua}},
  \bibinfo {author} {\bibfnamefont {X.}~\bibnamefont {Zambrana-Puyalto}},
  \bibinfo {author} {\bibfnamefont {X.}~\bibnamefont {Vidal}}, \bibinfo
  {author} {\bibfnamefont {G.}~\bibnamefont {Molina-Terriza}}, \ and\ \bibinfo
  {author} {\bibfnamefont {J.~J.}\ \bibnamefont {S\'aenz}},\ }\href {\doibase
  10.1103/PhysRevLett.114.113902} {\bibfield  {journal} {\bibinfo  {journal}
  {Phys. Rev. Lett.}\ }\textbf {\bibinfo {volume} {114}},\ \bibinfo {pages}
  {113902} (\bibinfo {year} {2015})}\BibitemShut {NoStop}%
\bibitem [{\citenamefont {Khanikaev}\ \emph {et~al.}(2013)\citenamefont
  {Khanikaev}, \citenamefont {Hossein~Mousavi}, \citenamefont {Tse},
  \citenamefont {Kargarian}, \citenamefont {MacDonald},\ and\ \citenamefont
  {Shvets}}]{Khanikaev2013}%
  \BibitemOpen
  \bibfield  {author} {\bibinfo {author} {\bibfnamefont {A.~B.}\ \bibnamefont
  {Khanikaev}}, \bibinfo {author} {\bibfnamefont {S.}~\bibnamefont
  {Hossein~Mousavi}}, \bibinfo {author} {\bibfnamefont {W.-K.}\ \bibnamefont
  {Tse}}, \bibinfo {author} {\bibfnamefont {M.}~\bibnamefont {Kargarian}},
  \bibinfo {author} {\bibfnamefont {A.~H.}\ \bibnamefont {MacDonald}}, \ and\
  \bibinfo {author} {\bibfnamefont {G.}~\bibnamefont {Shvets}},\ }\href
  {\doibase 10.1038/nmat3520} {\bibfield  {journal} {\bibinfo  {journal} {Nat
  Mater}\ }\textbf {\bibinfo {volume} {12}},\ \bibinfo {pages} {233} (\bibinfo
  {year} {2013})}\BibitemShut {NoStop}%
\bibitem [{\citenamefont {Silveirinha}(2017)}]{Silveirinha2017}%
  \BibitemOpen
  \bibfield  {author} {\bibinfo {author} {\bibfnamefont {M.~G.}\ \bibnamefont
  {Silveirinha}},\ }\href {\doibase 10.1103/PhysRevB.95.035153} {\bibfield
  {journal} {\bibinfo  {journal} {Phys. Rev. B}\ }\textbf {\bibinfo {volume}
  {95}},\ \bibinfo {pages} {035153} (\bibinfo {year} {2017})}\BibitemShut
  {NoStop}%
\bibitem [{\citenamefont {Lindell}\ \emph {et~al.}(2009)\citenamefont
  {Lindell}, \citenamefont {Sihvola}, \citenamefont {Yla-Oijala},\ and\
  \citenamefont {Wallen}}]{Lindell2009}%
  \BibitemOpen
  \bibfield  {author} {\bibinfo {author} {\bibfnamefont {I.~V.}\ \bibnamefont
  {Lindell}}, \bibinfo {author} {\bibfnamefont {A.}~\bibnamefont {Sihvola}},
  \bibinfo {author} {\bibfnamefont {P.}~\bibnamefont {Yla-Oijala}}, \ and\
  \bibinfo {author} {\bibfnamefont {H.}~\bibnamefont {Wallen}},\ }\href
  {\doibase 10.1109/TAP.2009.2027180} {\bibfield  {journal} {\bibinfo
  {journal} {IEEE Trans. Antennas Propagat.}\ }\textbf {\bibinfo {volume}
  {57}},\ \bibinfo {pages} {2725} (\bibinfo {year} {2009})}\BibitemShut
  {NoStop}%
\bibitem [{\citenamefont {Abdelrahman}\ \emph {et~al.}(2017)\citenamefont
  {Abdelrahman}, \citenamefont {Rockstuhl},\ and\ \citenamefont
  {Fernandez-Corbaton}}]{Abdelrahman2017}%
  \BibitemOpen
  \bibfield  {author} {\bibinfo {author} {\bibfnamefont {I.}~\bibnamefont
  {Abdelrahman}}, \bibinfo {author} {\bibfnamefont {C.}~\bibnamefont
  {Rockstuhl}}, \ and\ \bibinfo {author} {\bibfnamefont {I.}~\bibnamefont
  {Fernandez-Corbaton}},\ }\href@noop {} {\bibfield  {journal} {\bibinfo
  {journal} {Accepted, Sci. Rep.}\ } (\bibinfo {year} {2017})}\BibitemShut
  {NoStop}%
\bibitem [{\citenamefont {Zambrana-Puyalto}\ \emph
  {et~al.}(2013{\natexlab{b}})\citenamefont {Zambrana-Puyalto}, \citenamefont
  {Vidal}, \citenamefont {Juan},\ and\ \citenamefont
  {Molina-Terriza}}]{Zambrana2013b}%
  \BibitemOpen
  \bibfield  {author} {\bibinfo {author} {\bibfnamefont {X.}~\bibnamefont
  {Zambrana-Puyalto}}, \bibinfo {author} {\bibfnamefont {X.}~\bibnamefont
  {Vidal}}, \bibinfo {author} {\bibfnamefont {M.~L.}\ \bibnamefont {Juan}}, \
  and\ \bibinfo {author} {\bibfnamefont {G.}~\bibnamefont {Molina-Terriza}},\
  }\href@noop {} {\bibfield  {journal} {\bibinfo  {journal} {Opt. Express}\
  }\textbf {\bibinfo {volume} {21}},\ \bibinfo {pages} {17520} (\bibinfo {year}
  {2013}{\natexlab{b}})}\BibitemShut {NoStop}%
\bibitem [{\citenamefont {Mishchenko}\ \emph {et~al.}(2016)\citenamefont
  {Mishchenko}, \citenamefont {Zakharova}, \citenamefont {Khlebtsov},
  \citenamefont {Videen},\ and\ \citenamefont {Wriedt}}]{Mishchenko2016}%
  \BibitemOpen
  \bibfield  {author} {\bibinfo {author} {\bibfnamefont {M.~I.}\ \bibnamefont
  {Mishchenko}}, \bibinfo {author} {\bibfnamefont {N.~T.}\ \bibnamefont
  {Zakharova}}, \bibinfo {author} {\bibfnamefont {N.~G.}\ \bibnamefont
  {Khlebtsov}}, \bibinfo {author} {\bibfnamefont {G.}~\bibnamefont {Videen}}, \
  and\ \bibinfo {author} {\bibfnamefont {T.}~\bibnamefont {Wriedt}},\ }\href
  {\doibase http://dx.doi.org/10.1016/j.jqsrt.2015.11.005} {\bibfield
  {journal} {\bibinfo  {journal} {J. Quant. Spectrosc. Radiat. Transfer}\
  }\textbf {\bibinfo {volume} {178}},\ \bibinfo {pages} {276 } (\bibinfo {year}
  {2016})}\BibitemShut {NoStop}%
\bibitem [{\citenamefont {Rahimzadegan}\ \emph {et~al.}(2017)\citenamefont
  {Rahimzadegan}, \citenamefont {Alaee}, \citenamefont {Fernandez-Corbaton},\
  and\ \citenamefont {Rockstuhl}}]{rahimzadegan2017Fundamental}%
  \BibitemOpen
  \bibfield  {author} {\bibinfo {author} {\bibfnamefont {A.}~\bibnamefont
  {Rahimzadegan}}, \bibinfo {author} {\bibfnamefont {R.}~\bibnamefont {Alaee}},
  \bibinfo {author} {\bibfnamefont {I.}~\bibnamefont {Fernandez-Corbaton}}, \
  and\ \bibinfo {author} {\bibfnamefont {C.}~\bibnamefont {Rockstuhl}},\ }\href
  {\doibase 10.1103/PhysRevB.95.035106} {\bibfield  {journal} {\bibinfo
  {journal} {Phys. Rev. B}\ }\textbf {\bibinfo {volume} {95}},\ \bibinfo
  {pages} {035106} (\bibinfo {year} {2017})}\BibitemShut {NoStop}%
\bibitem [{\citenamefont {Ruan}\ and\ \citenamefont
  {Fan}(2010)}]{ruan2010superscattering}%
  \BibitemOpen
  \bibfield  {author} {\bibinfo {author} {\bibfnamefont {Z.}~\bibnamefont
  {Ruan}}\ and\ \bibinfo {author} {\bibfnamefont {S.}~\bibnamefont {Fan}},\
  }\href@noop {} {\bibfield  {journal} {\bibinfo  {journal} {Phys. Rev. Lett.}\
  }\textbf {\bibinfo {volume} {105}},\ \bibinfo {pages} {013901} (\bibinfo
  {year} {2010})}\BibitemShut {NoStop}%
\bibitem [{\citenamefont {Tojo}\ \emph {et~al.}(2004)\citenamefont {Tojo},
  \citenamefont {Hasuo},\ and\ \citenamefont {Fujimoto}}]{Tojo2004}%
  \BibitemOpen
  \bibfield  {author} {\bibinfo {author} {\bibfnamefont {S.}~\bibnamefont
  {Tojo}}, \bibinfo {author} {\bibfnamefont {M.}~\bibnamefont {Hasuo}}, \ and\
  \bibinfo {author} {\bibfnamefont {T.}~\bibnamefont {Fujimoto}},\ }\href@noop
  {} {\bibfield  {journal} {\bibinfo  {journal} {Phys. Rev. Lett.}\ }\textbf
  {\bibinfo {volume} {92}},\ \bibinfo {pages} {053001} (\bibinfo {year}
  {2004})}\BibitemShut {NoStop}%
\bibitem [{\citenamefont {Tojo}\ and\ \citenamefont {Hasuo}(2005)}]{Tojo2005}%
  \BibitemOpen
  \bibfield  {author} {\bibinfo {author} {\bibfnamefont {S.}~\bibnamefont
  {Tojo}}\ and\ \bibinfo {author} {\bibfnamefont {M.}~\bibnamefont {Hasuo}},\
  }\href@noop {} {\bibfield  {journal} {\bibinfo  {journal} {Phys. Rev. A}\
  }\textbf {\bibinfo {volume} {71}},\ \bibinfo {pages} {012508} (\bibinfo
  {year} {2005})}\BibitemShut {NoStop}%
\bibitem [{\citenamefont {Fernandez-Corbaton}\ \emph
  {et~al.}(2016{\natexlab{b}})\citenamefont {Fernandez-Corbaton}, \citenamefont
  {Zambrana-Puyalto}, \citenamefont {Bonod},\ and\ \citenamefont
  {Rockstuhl}}]{FerCor2016b}%
  \BibitemOpen
  \bibfield  {author} {\bibinfo {author} {\bibfnamefont {I.}~\bibnamefont
  {Fernandez-Corbaton}}, \bibinfo {author} {\bibfnamefont {X.}~\bibnamefont
  {Zambrana-Puyalto}}, \bibinfo {author} {\bibfnamefont {N.}~\bibnamefont
  {Bonod}}, \ and\ \bibinfo {author} {\bibfnamefont {C.}~\bibnamefont
  {Rockstuhl}},\ }\href {\doibase 10.1103/PhysRevA.94.053822} {\bibfield
  {journal} {\bibinfo  {journal} {Phys. Rev. A}\ }\textbf {\bibinfo {volume}
  {94}},\ \bibinfo {pages} {053822} (\bibinfo {year}
  {2016}{\natexlab{b}})}\BibitemShut {NoStop}%
\bibitem [{\citenamefont {Malassis}\ \emph {et~al.}(2013)\citenamefont
  {Malassis}, \citenamefont {Mass\'e}, \citenamefont {Tr\'eguer-Delapierre},
  \citenamefont {Mornet}, \citenamefont {Weisbecker}, \citenamefont {Kravets},
  \citenamefont {Grigorenko},\ and\ \citenamefont {Barois}}]{Malassis2013}%
  \BibitemOpen
  \bibfield  {author} {\bibinfo {author} {\bibfnamefont {L.}~\bibnamefont
  {Malassis}}, \bibinfo {author} {\bibfnamefont {P.}~\bibnamefont {Mass\'e}},
  \bibinfo {author} {\bibfnamefont {M.}~\bibnamefont {Tr\'eguer-Delapierre}},
  \bibinfo {author} {\bibfnamefont {S.}~\bibnamefont {Mornet}}, \bibinfo
  {author} {\bibfnamefont {P.}~\bibnamefont {Weisbecker}}, \bibinfo {author}
  {\bibfnamefont {V.}~\bibnamefont {Kravets}}, \bibinfo {author} {\bibfnamefont
  {A.}~\bibnamefont {Grigorenko}}, \ and\ \bibinfo {author} {\bibfnamefont
  {P.}~\bibnamefont {Barois}},\ }\href@noop {} {\bibfield  {journal} {\bibinfo
  {journal} {Langmuir}\ }\textbf {\bibinfo {volume} {29}},\ \bibinfo {pages}
  {1551} (\bibinfo {year} {2013})}\BibitemShut {NoStop}%
\bibitem [{\citenamefont {Ponsinet}\ \emph {et~al.}(2015)\citenamefont
  {Ponsinet}, \citenamefont {Barois}, \citenamefont {Gali}, \citenamefont
  {Richetti}, \citenamefont {Salmon}, \citenamefont {Vallecchi}, \citenamefont
  {Albani}, \citenamefont {Le~Beulze}, \citenamefont {Gomez-Grana},
  \citenamefont {Duguet}, \citenamefont {Mornet},\ and\ \citenamefont
  {Treguer-Delapierre}}]{Ponsinet2014}%
  \BibitemOpen
  \bibfield  {author} {\bibinfo {author} {\bibfnamefont {V.}~\bibnamefont
  {Ponsinet}}, \bibinfo {author} {\bibfnamefont {P.}~\bibnamefont {Barois}},
  \bibinfo {author} {\bibfnamefont {S.~M.}\ \bibnamefont {Gali}}, \bibinfo
  {author} {\bibfnamefont {P.}~\bibnamefont {Richetti}}, \bibinfo {author}
  {\bibfnamefont {J.~B.}\ \bibnamefont {Salmon}}, \bibinfo {author}
  {\bibfnamefont {A.}~\bibnamefont {Vallecchi}}, \bibinfo {author}
  {\bibfnamefont {M.}~\bibnamefont {Albani}}, \bibinfo {author} {\bibfnamefont
  {A.}~\bibnamefont {Le~Beulze}}, \bibinfo {author} {\bibfnamefont
  {S.}~\bibnamefont {Gomez-Grana}}, \bibinfo {author} {\bibfnamefont
  {E.}~\bibnamefont {Duguet}}, \bibinfo {author} {\bibfnamefont
  {S.}~\bibnamefont {Mornet}}, \ and\ \bibinfo {author} {\bibfnamefont
  {M.}~\bibnamefont {Treguer-Delapierre}},\ }\href {\doibase
  10.1103/PhysRevB.92.220414} {\bibfield  {journal} {\bibinfo  {journal} {Phys.
  Rev. B}\ }\textbf {\bibinfo {volume} {92}},\ \bibinfo {pages} {220414}
  (\bibinfo {year} {2015})}\BibitemShut {NoStop}%
\bibitem [{\citenamefont {Lee}\ and\ \citenamefont
  {El-Sharkawi}(2008)}]{lee2008modern}%
  \BibitemOpen
  \bibfield  {author} {\bibinfo {author} {\bibfnamefont {K.~Y.}\ \bibnamefont
  {Lee}}\ and\ \bibinfo {author} {\bibfnamefont {M.~A.}\ \bibnamefont
  {El-Sharkawi}},\ }\href@noop {} {\emph {\bibinfo {title} {Modern Heuristic
  Optimization Techniques}}},\ Vol.~\bibinfo {volume} {39}\ (\bibinfo
  {publisher} {John Wiley \& Sons},\ \bibinfo {year} {2008})\BibitemShut
  {NoStop}%
\bibitem [{\citenamefont {Kosters}\ \emph {et~al.}(2017)\citenamefont
  {Kosters}, \citenamefont {de~Hoogh}, \citenamefont {Zeijlemaker},
  \citenamefont {Acar}, \citenamefont {Rotenberg},\ and\ \citenamefont
  {Kuipers}}]{Kosters2017}%
  \BibitemOpen
  \bibfield  {author} {\bibinfo {author} {\bibfnamefont {D.}~\bibnamefont
  {Kosters}}, \bibinfo {author} {\bibfnamefont {A.}~\bibnamefont {de~Hoogh}},
  \bibinfo {author} {\bibfnamefont {H.}~\bibnamefont {Zeijlemaker}}, \bibinfo
  {author} {\bibfnamefont {H.}~\bibnamefont {Acar}}, \bibinfo {author}
  {\bibfnamefont {N.}~\bibnamefont {Rotenberg}}, \ and\ \bibinfo {author}
  {\bibfnamefont {L.}~\bibnamefont {Kuipers}},\ }\href {\doibase
  10.1021/acsphotonics.7b00496} {\bibfield  {journal} {\bibinfo  {journal} {ACS
  Photonics}\ }\textbf {\bibinfo {volume} {4}},\ \bibinfo {pages} {1858}
  (\bibinfo {year} {2017})}\BibitemShut {NoStop}%
\end{thebibliography}
\end{document}